\documentstyle[preprint,revtex]{aps} 
\begin{document}


\preprint
{
\begin{flushright}
LAEFF--93/015\\
November 1993
\end{flushright}
}

\begin{title}
Quark Flavors as Entropy Ordered States of QCD.\\
\end{title}

\author{J.\ P\'erez--Mercader\cite{AAAuth}}
\begin{instit}

Laboratorio de Astrof\'{\i}sica Espacial y F\'{\i}sica
Fundamental\\
Apartado 50727\\
28080 Madrid
\end{instit}

\vspace{-.125in}
\begin{center} \bf
{\sl Submitted to Physical Review Letters}

\end{center}

\begin{abstract}

We discuss a natural notion of entropy in quantum field theory
and apply it to asymptotically free theories in their
perturbative regimes. We then specialize to QCD and find that quark
flavor states can be described as entropy--ordered states of QCD,
and that the masses for the $s\bar{s}$--state, charm,
$c\bar{c}$--state, bottom and $b\bar{b}$--state can all be fitted
by requiring that the entropy of each of these states be the same.
The resulting Pearson correlation coefficient between theory and
experiment is better than 0.99, and the known quark masses can be
accounted for with less than an 8\% error.
\end{abstract}



The notion of entropy is very general \cite{lebowitz}: given a
probability distribution one can calculate the entropy associated
with it and derive information about the collective behavior of
the physical system represented by the distribution.

A general feature of quantum theory is the existence of
quantum fluctuations. They modify the physical quantities into
$scale$ dependent objects which reflect the quantum nature of the
underlying virtual cloud: at a given scale, where quantum
fluctuations are active, there are contributions from virtual
processes involving momentum transfers which are smaller or equal
than what the uncertainty principle dictates for the size of the
probed region. In quantum field theory, the corrections are
taken into account via the renormalization process, and their
scale dependence described by appropriate renormalization group
equations \cite{gellmannlow} (RGEs) satisfied by the $n$--point
functions.

We will extend the notion of entropy to quantum field theory by
taking advantage of the succinct description of quantum phenomena
given by the renormalization group. Then we will apply it to QCD
in the leading approximation.

At 1--loop, for massless mediating quanta (such as gluons or
photons) and at distances close to the Compton wavelength of some
degree of freedom, the $leading$ term of the static interaction
energy between two elementary charges is easily seen to be given by
\cite{self}

\begin{equation}
V(r) \cong g_0^2 r^{-1+\sigma}
\label{1}
\end{equation}

\noindent
where, at one loop, the RGE for \footnote{Here $\mu$
is an arbitrary momentum scale. For notation and references cf.
Ref. \cite{rothe}.} $g_0^2$, $\mu d g_0^2/d \mu =-\beta_0 g_0^4$,
determines $\sigma$ as $\sigma = +2 \beta_0g_0^2$. Because of the
Poisson equation satisfied by Eq.(\ref{1}) and the
well studied connection between potential theory and probability
theory \cite{doob}, one can associate with $V(r)$ a density
$\rho (r)$ and $interpret$ $\rho$ as a probability (in the
classical sense of probability) density for the
distribution of the virtual cloud that surrounds the
elementary charges in the quantum mechanical vacuum. Away from
$r=0$, $\rho (r)$ turns out to be

\begin{equation}
\rho (r) =A r^{-3+\sigma}
\label{2}
\end{equation}

\noindent
with  $A={\sigma  \over {4\pi}}R_0^{-\sigma }$ when $\sigma >0$
and $A={{-\sigma }\over {4\pi }}r_0^{+\sigma}$ if $\sigma < 0$.
The quantities $R_0$ and $r_0$ are, respectively, effective $IR$
and $UV$ cutoffs necessary to ensure the normalizability of the
probability distribution in the asymptotically free (AF) and
non--asymptotically free (NAF) cases.

The fine--grained entropy for the probability density of Eq.
(\ref{2}) can be directly computed using its definition

\cite{lebowitz}, or one can perform a simple variational
calculation \cite{montroll} and get the (fine grained)
Gibb's entropy $S= -\int \rho \ln \rho d^3
\overrightarrow{r}$, identify\footnote{We set the ``Boltzmann
constant" equal to unity.} a ``temperature" $T=(3-\sigma)^{-1}$,
and an internal energy $U$.

We now restrict ourselves to AF--theories, where the entropy is
given by

$$
-S^{(\sigma > 0)}= \int_{\Omega (R_0)} \rho (r) \ln [\Theta
\rho(r)] d^3 \overrightarrow{r}

$$
$$
=\left(1-\left(\frac{r}{R_0}\right)^{\sigma}\right)
\left(\frac{3}{\sigma} -1 + \ln \frac{\sigma}{4 \pi}\right)
+(3-\sigma) \left(\frac{r}{R_0}\right)^{\sigma} \ln r

$$
\begin{equation}
+\sigma \left( \frac{r}{R_0}\right)^\sigma \ln R_0
-3 \ln R_0 +

\left(1-\left(\frac{r}{R_0}\right)^{\sigma}\right) \ln \Theta
\label{entropy}
\end{equation}

Borrowing from thermodynamics \cite{fermi}, the region of
integration has been constrained to the one contained within the
ball of radius $R_0$; this determines the limiting state
accessible to the quantum field system in the $IR$. \footnote{At
least from the point of view of canonical perturbation theory.}
The lower limit in the integral, $r$, can be arbitrarily small
since (for AF--theories) there are no singularities as $r$ goes to
zero. Also, to make the argument of the logarithm
dimensionless, it is necessary to introduce a dimensional constant
$\Theta$ which is a reflection of Nernst theorem in statistical
thermodynamics. We take $\Theta= \bar{r}^3 \exp a$, where $\bar{r}
< R_0$ and, at this stage, $a$ is an undetermined quantity.

Taking the lower limit to be $r=0$, and choosing $\bar{r}$ as the
$maximum$ system size supported at a given value of $\sigma$, we
immediately get the following expression for the entropy of the
system

\begin{equation}
S^{(\sigma > 0)}=1-\frac{3}{\sigma} -\ln \frac{\sigma}{4\pi}
+ 3 \ln \frac{R_0}{\bar{r}} -a
\label{3}
\end{equation}

\noindent
 From Eq. (\ref{3}) we see that the system is
thermodynamically stable when $\sigma < 3$ \cite{reichl}.

We now apply the above ideas to the
quark sector of QCD, where

$$
\beta_0 =\frac{1}{16 \pi^2}\left( 11- \frac{2}{3}n_f\right)
$$

\noindent
Here $n_f$ is the number of flavors excited from the vacuum at the
momentum scale $\mu$. As we
increase $\mu$, $n_f$ increases and $\beta_0$ becomes less
positive; at the same time, and because of the asymptotic freedom
of QCD, $g_0^2$ also decreases; in other words, $\sigma$ becomes
less positive and, cf. Figure 1, the ``temperature" of
our quantum field system is lowered.

We can set the zero of the entropy at any scale, i.e., at any value
of $\sigma$. This fixes the constant $a$. We choose the infrared
cutoff $R_0$ to be the size\footnote{Which we
take to be the largest system size at which one could begin to
``trust" perturbative QCD.} of the strange
quark--antiquark system, and the value of $a$ as\footnote{The

constant $k$ will be used later to fit the experimental
data.}

\begin{equation}
a= 1- \frac{3}{\sigma(2m_{strange})}-
\ln\frac{\sigma(2m_{strange})}{4\pi} + k
\label{4}
\end{equation}

The ``temperature" and the entropy as functions of $\sigma$ are
shown in Figure 1. For $\sigma <3$, when we probe the
system at shorter and shorter distances, the system ``cools", and
the entropy decreases. In other words, going into the $UV$ is
accompanied by a drop in the temperature which, from the point of
view of order--disorder, is a ``disorder--decreasing" process

reflected in QCD by a decrease in entropy. But in quantum field
theory as we decrease the size of the system being probed, more
momentum transfer is available,  $new$ particle thresholds are
crossed and $more$ degrees of freedom become active in the quantum
system; this contribution of additional matter degrees of freedom
at shorter distances raises the number of states accessible to
the quantum system, which is a ``disorder--increasing" process
and thus, makes the entropy grow.

The $simplest$ possibility for the net entropy is to $assume$ that
these two competing effects compensate each other, and that the
process of going into the ultraviolet is isentropic. We will
assume this in what follows.

The adiabatic nature of the process implies that the entropy stays
constant as we go deeper into the $UV$, and Eq. (\ref{3})
with $\Delta S=0$ leads to an equation relating $\sigma$ at some
scale $\bar{r}$ with $\ln(R_0/\bar{r})$. This means that the
appearance of a new flavor in $\beta_0$ can $not$ occur at an
arbitrary scale: only those values consistent with $\Delta S=0$
are allowed. That is, at scales $\bar{r}_i$ and $\bar{r}_{i+1}$
related by

$$
3\left(1-\frac{\beta_0(i)}{\beta_0(i+1)}\right) y_c
+\ln \left[1+2\beta_0(i) g_0^2(\bar{r}_i)y_c\right]
$$
\begin{equation}
=\ln \frac{\beta_0(i+1)}{\beta_0(i)}+\frac{3}{2g_0^2(\bar{r}_i)}
\cdot
\left[ \frac{1}{\beta_0(i+1)}-\frac{1}{\beta_0(i)} \right]
\label{5}
\end{equation}

\noindent
Here $i$ is the number of flavors at the (length) scale
$\bar{r}_i$ and $y_c \equiv \ln \bar{r}_i/\bar{r}_{i+1}$.
This equation shows that $unless$ $\beta_0(i) \neq
\beta_0(i+1)$, $y_c \equiv 0$. It can also be easily seen from
Eq. (\ref{5}) that $i$ can not grow indefinitely: there is a
maximum value of $i$ beyond which there is no solution for $y_c$.
In other words: it implies the existence of an {\it upper} bound to
the number of flavors.

In Figure 2, we show one\footnote{Due to reasons of space, the
upper bound on $n_f$ and the masses of the unknown flavors,
including the predicted value for the mass of the $top$, are
discussed in a separate publication \cite{forthcoming}.} of the
consequences of the ideas just discussed. We compare the
known experimental heavier quark masses (i.e., all except $up$ and
$down$) and strong interaction coupling constant, with the
$result$ of $assuming$ $\Delta S=0$ between quark
states. The experimental data can be described
by adjusting $one$ single free parameter, the zero of the entropy:
the resulting correlation between theory and experiment is better
than .99. For Figure 2 we have fitted Eq.(\ref{3}), with $k$ of
Eq.(\ref{4}), to the known experimental data \cite{rpp}: {\it the
product is that the theoretical curve is never more than 8\% away
 from the central experimental values}. This supports the
interpretation of quark flavors as entropy--driven,
self--organized \cite{nature} states of QCD, and the scenario
described above for the transition into the $UV$ as an isentropic
process, where the disorder (entropy increasing) increasing
process of probing the QCD system at shorter and shorter distances
is compensated by the order (entropy decreasing) increasing
process of lowering the ``temperature". Quantum fluctuations are
the messengers in this phylogeny of quarks, since they are
responsible for both: making available more degrees of freedom
 from the vacuum $and$ for generating the scaling corrections which
lead to our probability densities.

\acknowledgements

I have benefitted from discussions with M. Bastero and T. Goldman.

\figure{The ``temperature" $T$ and entropy $S$ plotted against
$\sigma$ in AF--theories. The zero of the entropy has been
(arbitrarily) chosen so as to display the curves in a position
where their features can be best appreciated. We also show the
maximum of the entropy occurring at $\sigma =3$.}

\figure{The condition $\Delta S =0$ for quarks in QCD. The
experimental data have been fitted setting $S^{(\sigma>0)}
(\sigma)$ of Equation (\ref{3}) equal to zero, with

(cf. Eq. (\ref{4})) $k=-0.176136$. The experimental values are
given in Table 1. The Pearson's correlation coefficient between
experimental data and theory is greater than 0.99.
In the inset we plot the residues for this fit. The reduced
$\chi^2$ is 0.1120. All known quark masses are fitted with an error
of less than 8\% .}

\newpage

\begin{table}
\caption{Experimental values used in fit.$^{\rm a}$}
\begin{tabular}{c|c|c|c}
Flavor/State & $<3\ \ln (m_f/2 m_{strange})>$ & $\sigma_{exp}$ &
Residue\\

\tableline
$b\bar{b}$ & 8.440232  & 0.206233 & -0.27553 \\
$b$ & 6.360791  & 0.240759 & -0.42370 \\
$c\bar{c}$ & 4.828314  & 0.302973 & 0.37270 \\
$c$ & 2.748872  & 0.384270 & 0.15042 \\
$s\bar{s}$ & 0.0  & 0.670211 & 0.17611 \\
\end{tabular}
\tablenotes{$^{\rm a}$ $\alpha_{strong} (M_Z)= 0.116 \pm 0.005$.
The scale for $2 m_{strange}$ was taken at 0.6 $GeV$; the
data on quark masses is from Ref. \cite{rpp}.} \end{table}
\end{document}